\documentclass[floatfix,showpacs,superscriptaddress,prl,twocolumn]{revtex4}%
\usepackage{amsbsy,amssymb,amsmath,bm}
\usepackage{graphicx,color}
\usepackage{amsfonts}
\usepackage{amssymb}
\usepackage{amsmath}
\usepackage{graphicx}%
\ifx\pdfoutput\relax\let\pdfoutput=\undefined\fi
\newcount\msipdfoutput
\ifx\pdfoutput\undefined\else
\ifcase\pdfoutput\else
\ifx\paperwidth\undefined\else
\ifdim\paperheight=0pt\relax\else\pdfpageheight\paperheight\fi
\ifdim\paperwidth=0pt\relax\else\pdfpagewidth\paperwidth\fi
\fi\fi\fi
\begin{document}
\title{Ballistic transport in graphene beyond linear response \bigskip}
\author{B. Rosenstein }
\email{vortexbar@yahoo.com}
\affiliation{\textit{Electrophysics Department, National Chiao Tung University, Hsinchu
30050,} \textit{Taiwan, R. O. C}.}
\author{M. Lewkowicz}
\affiliation{\textit{Applied Physics Department, Ariel University Center of Samaria, Ariel
40700, Israel}}
\author{H.C. Kao}
\affiliation{\textit{Physics Department, National Taiwan Normal University, Taipei 11650,
Taiwan, R. O. C}.}
\author{Y. Korniyenko }
\affiliation{\textit{Electrophysics Department, National Chiao Tung University, Hsinchu
30050,} \textit{Taiwan, R. O. C}.}
\date{\today }

\begin{abstract}
The process of coherent creation of particle - hole excitations by an electric
field in graphene is quantitatively described beyond linear response. We
calculate the evolution of current density, number of pairs and energy in
ballistic regime for electric field $E$ using the tight binding model. While
for ballistic flight times smaller than $t_{nl}\varpropto E^{-1/2}$ current is
linear in $E$ and independent of time, for larger ballistic times the current
increases after $t_{nl}$ as $J\varpropto E^{3/2}t $ and finally at yet larger
times ($t>t_{B}\varpropto E^{-1}$) Bloch oscillations set in. It is shown that
the number of pairs follows the 2D generalization of the Schwinger's creation
rate $n\varpropto E^{3/2}$ only on certain time segments with a prefactor
different from that obtained using the asymptotic formula.

\end{abstract}

\maketitle

\section{Introduction}

It became increasingly evident that electronic mobility in graphene is
extremely large exceeding that in best semiconductor 2D samples
\cite{GeimPRL08}. Since the system is so clean the transport becomes
ballistic, especially in suspended graphene samples \cite{Andrei08}, so that
interactions of electrons with phonons, ripplons, disorder and among
themselves can be neglected. Therefore there is a chance to observe various
theoretically predicted exotic phenomena like nonlinear response and Bloch
oscillations \cite{Davis}.

Generally ballistic transport occurs due to two distinct phenomena. If a
mobile charge carrier is available (as in an electron gas in a metal),
electric field accelerates it, so that current increases linearly in time. In
addition, an electric field can create new charge carriers (the process
typically suppressed by energy gaps). A peculiarity of the ballistic transport
in graphene with the Fermi level pinned right on the two Dirac points
\cite{Castro} (which happens naturally in suspended graphene samples
\cite{Andrei08}) is that there are no charge carriers present at all. The
Fermi surface therefore shrinks to just two points. The carriers are created
solely by an applied electric field like in the Zener tunneling effect in
semiconductors \cite{Davis}, but energy gap vanishes due to "ultra -
relativistic" dispersion relation, $\varepsilon=v_{g}\left\vert \mathbf{k}%
\right\vert ,$ where $v_{g}\sim10^{6}m/\sec$ is the graphene velocity.

The electron - hole pairs are created fast enough to make the current linear
in electric field and constant in time, so that it looks like a Drude type
linear response due to disorder rather than the ballistic acceleration
$J\varpropto t$ of an electron gas with finite carrier density. The illusion
of the "Ohmic" behaviour however cannot continue indefinitely in the absence
of scatterers, and should eventually cross over to some sort of "acceleration"
or even Bloch oscillations at large times. The behaviour is expected to become
nonlinear as function of electric field as indicated by the nonlinearity of
the pair creation rate. It was shown long time ago \cite{Schwinger}, in the
context of particle physics, that the pair creation rate at asymptotically
large times is proportional to $E^{3/2}$.

Ambiguities in the application of the standard Kubo approach for the ultra -
relativistic spectrum \cite{Ziegler}, led us propose a dynamic approach to the
tight binding model of graphene \cite{Lewkowicz}. Within leading order in $E$
(linear response) we found that the DC conductivity is $\sigma_{2}=\frac{\pi
}{2}\frac{e^{2}}{h}$ rather the often cited value $\frac{4}{\pi}\frac{e^{2}%
}{h}$ obtained both from Kubo formula \cite{Fradkin} and within the Landauer
formalism \cite{Katsnelson06}. In this note we solve the tight binding model
for arbitrary constant electric field. The evolution of current density
demonstrates that the crossover from the "Ohmic" regime to the nonlinear one
occurs at the experimentally achievable time scale $t_{nl}\propto E^{-1/2}$.
Bloch oscillations are shown to set in on scale $t_{B}\propto E^{-1}$ much
longer than $t_{nl}$ for experimentally accessible electric fields. We discuss
the relevance of the 2D generalization of the Schwinger's creation rate
formula \cite{Cohen} to physics of graphene.

\section{Tight binding model and its exact solution\textit{\ }}

Electrons in graphene are described sufficiently well for our purposes by the
2D tight binding model of nearest neighbor interactions on a honeycomb lattice
\cite{Castro}. The Hamiltonian in momentum space is%

\begin{equation}
\widehat{H}=\sum_{\mathbf{k}}%
\begin{pmatrix}
c_{\mathbf{k}}^{1\dag} & c_{\mathbf{k}}^{2\dag}%
\end{pmatrix}
H_{\mathbf{p}}%
\begin{pmatrix}
c_{\mathbf{k}}^{1}\\
c_{\mathbf{k}}^{2}%
\end{pmatrix}
;\text{ \ \ }H_{\mathbf{p}}=%
\begin{pmatrix}
0 & h_{\mathbf{p}}\\
h_{\mathbf{p}}^{\ast} & 0
\end{pmatrix}
, \label{H}%
\end{equation}
where
\begin{equation}
h_{\mathbf{p}}=-\gamma\left[  \exp\left(  i\frac{ap_{y}}{\sqrt{3}}\right)
+b\exp\left(  -i\frac{ap_{y}}{2\sqrt{3}}\right)  \right]  \label{h}%
\end{equation}
with $\gamma=2.7eV$ being the hopping energy and sum is over the Brillouin
zone. Nearest neighbors are separated by distance $a=3\mathring{A}%
,b=2\cos\left(  ak_{x}/2\right)  $ and pseudospin index denotes two triangular
sublattices. We consider the system in a constant and homogeneous electric
field $E$ along the $y$ direction switched on at $t=0$. It is described by the
minimal substitution $\mathbf{p}=\hslash\mathbf{k}+\frac{e}{c}\mathbf{A} $
with vector potential $\mathbf{A}=\left(  0,-cEt\right)  $ for $t>0$. Since
the crucial physical effect of the field is a coherent creation of electron -
hole pairs, mostly near the two Dirac points, a convenient formalism to
describe the pair creation is the "first quantized" formulation described in
detail in \cite{Gitman,Lewkowicz}. To consider the ballistic transport at zero
temperature, $T=0$ dynamically, one starts at time $t=0$ from the zero field
state in which all the negative energy one - particle states, $-\left\vert
h_{\mathbf{k}}\right\vert \equiv-\varepsilon_{\mathbf{k}}$, are occupied. The
second quantized state evolving from it\ is uniquely characterized by the
first quantized amplitude,
\begin{equation}
\psi_{\mathbf{k}}\left(  t\right)  =%
\begin{pmatrix}
\psi_{\mathbf{k}}^{1}\left(  t\right) \\
\psi_{\mathbf{k}}^{2}\left(  t\right)
\end{pmatrix}
\text{,} \label{spinor}%
\end{equation}
which is a "spinor" in the sublattice space. It obeys the matrix Schroedinger
equation
\begin{equation}
i\hslash\partial_{t}\psi_{\mathbf{k}}=H_{\mathbf{p}}\psi_{\mathbf{k}}\text{.}
\label{Schroedinger}%
\end{equation}

It is a peculiar property of the tight binding matrix Eq.(\ref{H}) that
solution for arbitrary $k_{y}$ can be reduced to that for $k_{y}=0$ and has
the Fourier series:%
\begin{align}
\psi_{\mathbf{k}}^{1}\left(  t\right)   &  =\sum_{s=\pm1}A^{s}\sum_{m=-\infty
}^{\infty}p_{m}^{s}\exp\left(  -i\omega_{m}^{s}\overline{t}\right)  ;\text{
\ }\label{Fourier}\\
\psi_{\mathbf{k}}^{2}\left(  t\right)   &  =-\sum_{s=\pm1}A^{s}\sum
_{m=-\infty}^{\infty}\frac{p_{m}^{s}+bp_{m-1}^{s}}{\omega_{m}^{-s}}\exp\left(
i\omega_{m}^{-s}\overline{t}\right)  \text{,}\nonumber
\end{align}
where $\overline{t}=t-t_{\gamma}ak_{y}/\mathcal{E}$ and $\omega_{m}^{s}%
=\omega^{s}+3\Omega m$ for frequency $\Omega=\mathcal{E}/\left(  2\sqrt
{3}t_{\gamma}\right)  ;$ $\mathcal{E}=E/E_{0}$. The relevant microscopic time
scale is $t_{\gamma}=\hslash/\gamma$ and field $E_{0}=\gamma/\left(
ea\right)  $. Recursion relations for the Fourier amplitudes $p_{m}$,%

\begin{align}
p_{m}  &  =\left[  \left(  \omega_{m}^{2}-8\Omega\omega_{m}+15\Omega
^{2}-1\right)  /b-b\left(  \omega_{m}-5\Omega\right)  \right]  p_{m-1}%
\nonumber\\
&  -\frac{\omega_{m}-2\Omega}{\omega_{m}-5\Omega}p_{m-2}\text{,} \label{rec}%
\end{align}
has two solutions $p^{s}$, $s=\pm1$ with two Floquet frequencies
\cite{Chicone} $\omega^{s}$. The recursion is easily solved numerically and
has the following convergent expansion in $b$ in the whole relevant range,
$0<b\leq2$,%
\begin{align}
\omega^{s}  &  =\omega_{0}^{s}+\frac{b^{2}}{\omega_{0}^{s2}-\Omega^{2}}\left[
\frac{\omega_{0}^{s}-2\Omega}{6\Omega\left(  2\omega_{0}^{s}+\Omega\right)
}+\frac{\omega_{0}^{s}}{2}-\Omega\right] \label{pert}\\
&  -\frac{b^{2}\left(  \omega_{0}^{s}-5\Omega\right)  }{6\Omega\left(
\omega_{0}^{s}-\Omega\right)  \left(  \omega_{0}^{s}-2\Omega\right)  \left(
2\omega_{0}^{s}-5\Omega\right)  }+O\left(  b^{4}\right)  ,\nonumber
\end{align}
with $\omega_{0}^{s}=s\Omega+\sqrt{t_{\gamma}^{-2}+\Omega^{2}}$. It turns out
that the two Floquet frequencies obey the relation obeying $\omega^{+}%
=2\Omega-\omega^{-}$, again peculiar to graphene, as can be checked by both
the perturbation theory, Eq.(\ref{pert}) and numerical results. For
experimentally accessible cases $\Omega<<t_{\gamma}^{-1}$ and the frequencies
are just $\pm t_{\gamma}^{-1}$. Coefficients $A^{s}$ are fixed by initial
conditions
\begin{equation}
\psi_{\mathbf{k}}\left(  t=0\right)  =u_{\mathbf{k}}=%
\begin{pmatrix}
1\\
-h_{\mathbf{k}}^{\ast}/\varepsilon_{\mathbf{k}}%
\end{pmatrix}
\text{.} \label{initial}%
\end{equation}
This solution is used to calculate evolution of current density, energy and
the number of electron - hole pairs.

\section{Time scale for observation of the Bloch oscillations in graphene}

Evolution of the current density during the ballistic "flight time" $t_{bal}$
is the integral over Brillouin zone (multiplied by factor $2$ due to spin)
\cite{Lewkowicz}:%

\begin{equation}
J_{y}\left(  t\right)  =-2e\sum_{\mathbf{k}}\psi_{\mathbf{k}}^{\dag}\left(
t\right)  \frac{\partial H_{\mathbf{p}}}{\partial p_{y}}\psi_{\mathbf{k}%
}\left(  t\right)  . \label{J}%
\end{equation}
The current density divided by electric field, $\sigma\left(  t\right)  \equiv
J_{y}\left(  t\right)  /E$, is shown in Fig.1 and 2 for various values of the
dimensionless electric field $\mathcal{E}$ in the range $2^{-8}-2^{-5} $.%

\begin{figure}
[ptb]
\begin{center}
\includegraphics[width=\columnwidth]{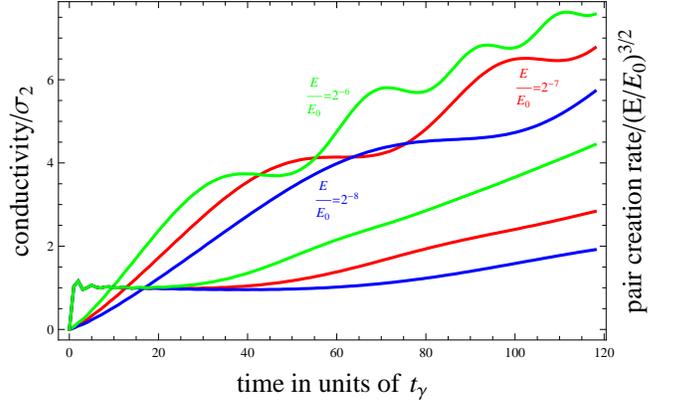}
\caption{The time evolution of the current density (3 bottomcurves) and the
scaled pair creation rate (3 top curves) at relatively short times for various
electric fields.}%
\label{Fig. 1}%
\end{center}
\end{figure}

Fig. 1 in which evolution is shown up to ballistic time of $120t_{\gamma},$
demonstrates that after an initial fast increase on the microscopic time scale
$t_{\gamma}$ (shown in more detail, using linear response, in \cite{Lewkowicz}%
), $\sigma\left(  t\right)  $ approaches the universal value $\sigma_{2}$ and
settles there. Beyond linear response one does not expect the current density
to hold up to this value indefinitely. In a ballistic system the energy
initially increases, as follows from the Joule law. The total energy of
electrons can be written in the first quantized formalism as
\begin{equation}
U_{tot}\left(  t\right)  =2\sum_{\mathbf{k}}\psi_{\mathbf{k}}^{\dag}\left(
t\right)  H_{\mathbf{p}}\psi_{\mathbf{k}}\left(  t\right)  \equiv2\left\langle
\psi\left(  t\right)  \left\vert H\right\vert \psi\left(  t\right)
\right\rangle \text{.} \label{energy}%
\end{equation}
It can be shown using Eq.(\ref{Schroedinger}) that the power%

\begin{align}
P\left(  t\right)   &  =\frac{d}{dt}U_{tot}=2\left\langle \psi\left\vert
\frac{d}{dt}H\right\vert \psi\right\rangle \label{power}\\
&  =-2eE\left\langle \psi\left\vert \frac{\partial H_{\mathbf{p}}}{\partial
p_{y}}\right\vert \psi\right\rangle =EJ_{y}\left(  t\right)  \text{,}\nonumber
\end{align}
is indeed proportional to current density. Since in the tight binding model
electron's energy cannot exceed the upper band edge energy $3\gamma$, hence at
some time scale $t_{B}$ the energy increase is reversed. The physics which
takes over is that of the Bloch oscillations and is similar to that in
ordinary materials, namely, electrons' energies are elevated by the electric
field \cite{Davis} due to the quasi-momentum shift. This feature is not
related to the unique "relativistic" feature of the graphene spectrum.%

\begin{figure}
[ptb]
\begin{center}
\includegraphics[width=\columnwidth]{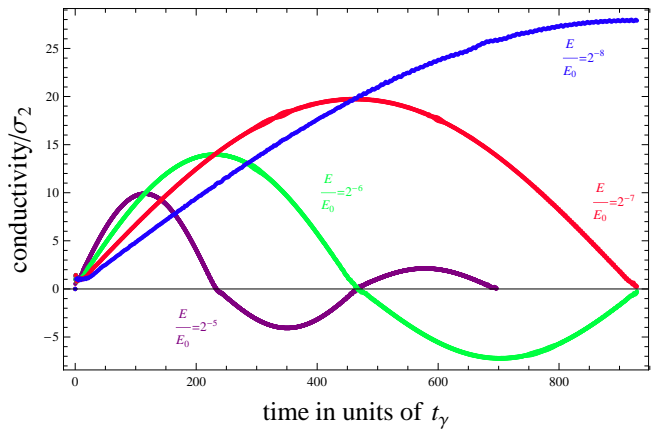}
\caption{The time evolution of the current density is shown for various fields
up to $1000$ $t_{\gamma}$}%
\label{Fig. 2}%
\end{center}
\end{figure}

The current, shown in Fig.2 for ballistic times up to $1000t_{\gamma}$, indeed
exhibits Bloch oscillations. It turns out that the current vanishes at points
given \textit{exactly }at multiples of $t_{B}/2$ with%

\begin{equation}
t_{B}=\frac{8\pi}{\sqrt{3}}\frac{\hbar}{eEa}=\frac{8\pi}{\sqrt{3}}%
\frac{t_{\gamma}}{\mathcal{E}} \label{t_Bloch}%
\end{equation}
being the period of the Bloch oscillations. The Bloch time is approximately
the time required for the electric field to shift the momentum across the
Brillouin zone $\Delta p_{y}=eEt_{B}\sim\hbar/a$. These times are very long
for experimentally achieved fields, much longer than the ballistic flight
time. One observes in Fig.2 another peculiar feature that (apart from the
"relativistic" initial constant segment) time dependence of $\sigma\left(
t\right)  $ is similar for different electric fields. Indeed, if one plots
$J/\sqrt{E}$ versus $tE$, all the curves nearly coincide. Moreover
\begin{equation}
J\left(  t\right)  =\sqrt{3}\sigma_{2}E_{0}^{1/2}E^{1/2}\sin\left(  \frac{2\pi
t}{t_{B}}\right)  \label{sin}%
\end{equation}
is an excellent fit.

For a sample of submicron dimensions, $L=0.5\mu m$, $W=1.5\mu m$, the
ballistic time can be estimated as $t_{bal}=L/v_{g}\simeq2.3\cdot
10^{3}t_{\gamma}$ with $v_{g}=\frac{\sqrt{3}}{2}\frac{\gamma a}{\hbar}$. For
current as large as $I_{\max}=mA$ the electric field is $E_{\max}%
=\frac{I_{\max}}{W\sigma_{2}}=10^{7}V/m\ $corresponding to $\mathcal{E}%
=10^{-3}$ (voltage in such case would be quite large $V_{\max}=5V$). The first
maximum of the Bloch oscillation will be seen at flight time of $t_{B}%
/4=3.6\cdot10^{3}t_{\gamma}$, which is of the same order as $t_{bal}$. If one
uses a value of the current typical to transport measurements $I=50\mu A$, the
electric field is just $E=5\cdot10^{5}V/m\ $corresponding $\mathcal{E}%
=5\cdot10^{-5}$, $t_{B}/4=7.2\cdot10^{4}t_{\gamma}>>t_{bal}$ and is therefore
out of reach.

\section{The crossover from linear to nonlinear regimes}

In Fig.1 one clearly observes a remarkable feature: there is a much smaller
crossover time $t_{nl}$, after which the conductivity rises linearly with time
above the constant "universal" value $\sigma_{2}$:
\begin{equation}
J\left(  t\right)  =\sigma_{2}\left(  \frac{\sqrt{3}}{2}E\right)
^{3/2}\left(  \frac{ev_{g}}{\hbar}\right)  ^{1/2}t\text{.} \label{J2}%
\end{equation}
The crossover time is
\begin{equation}
t_{nl}=\frac{2^{3/2}}{3^{3/4}}\sqrt{\frac{\hbar}{eEv_{g}}}\approx\frac
{1.3}{\sqrt{\mathcal{E}}}t_{\gamma}\text{.} \label{t_cros}%
\end{equation}
It becomes the same as the ballistic time $t_{bal}=2.3\cdot10^{3}t_{\gamma} $,
mentioned above, for relatively weak fields $E=10^{4}V/m$ corresponding to
$\mathcal{E}=10^{-6}$. Therefore some of the transport measurements performed
might be influenced by the physics beyond linear response.

A qualitative picture of this resistivity without dissipation is as follows.
The electric field creates electron - hole excitations mostly in the vicinity
of the Dirac points in which electrons behave as massless relativistic
fermions with the graphene velocity $v_{g}$ playing a role of velocity of
light. For such particles the absolute value of the velocity is $v_{g}$ and
cannot be altered by the electric field and is not related to the wave vector
$\mathbf{k}$\textbf{. }On the other hand, the orientation of the velocity is
influenced by the applied field.\textbf{\ }The electric current is
$e\mathbf{v}$, thus depending on orientation, so that its projection on the
field direction $y$ is increased by the field. The energy of the system
(calculated in a way similar to the current) is increasing continuously if no
channel for dissipation is included. Therefore the "Ohmic" conductivity
originates in creation of pairs near the Dirac points with an additional
contribution due to the alignment of the particles' motion with the field's
direction. At times of order $t_{B}$ his process exhausts itself due to the
following processes. Electrons gain momentum from the electric field and leave
eventually the neighborhoods of the Dirac points. They are no longer
ultra-relativistic and are described by (positive or negative) effective mass
\cite{Davis} and the more customary physics takes over.

The crossover to the nonlinear regime can be detected from within the
perturbation theory in electric field. Indeed we found that the $E^{2}$
correction to conductivity is
\begin{equation}
J\left(  t\right)  /E=\sigma_{2}\left[  1+\frac{3}{64}\mathcal{E}^{2}%
\frac{t^{4}}{t_{\gamma}^{4}}+O\left(  \mathcal{E}^{4}\right)  \right]  .
\label{J_correction}%
\end{equation}
The correction therefore becomes as large as the leading order for
$t=2.1t_{\gamma}/\mathcal{E}^{1/2}\simeq t_{nl}$. To gain more insight into
the nature of the crossover to nonlinear response we calculated also evolution
of the energy and number of electron - hole pairs during the ballistic flight.

\section{Schwinger's pair creation formula and graphene}

The states in the conduction band for each momentum $\mathbf{k}$ in the
Brillouin zone are described by a pseudospinor%
\begin{equation}
v_{\mathbf{k}}=%
\begin{pmatrix}
1\\
h_{\mathbf{k}}^{\ast}/\varepsilon_{\mathbf{k}}%
\end{pmatrix}
\label{v}%
\end{equation}
orthogonal to $u_{\mathbf{k}}$ defined in Eq.(\ref{initial}). The amplitude of
lifting of an electron into the conduction band is $A_{\mathbf{k}%
}=\left\langle \psi\left(  t\right)  |v_{\mathbf{k}}\right\rangle $ and
consequently the density of pairs (factor $2$ for spin) reads,%

\begin{equation}
N_{p}\left(  t\right)  =2\sum_{\mathbf{k}}\left\vert A_{\mathbf{k}}\right\vert
^{2}=2\sum_{\mathbf{k}}\left\vert \psi_{1}^{\ast}+\frac{h_{\mathbf{k}}^{\ast}%
}{\varepsilon_{\mathbf{k}}}\psi_{2}^{\ast}\right\vert ^{2}, \label{N_p}%
\end{equation}
and the rate $\frac{d}{dt}N_{p}$ is shown in Fig.1 as function of time. Its
time dependence exhibits several time scales. At times smaller than $t_{nl}$
expansion in electric field is applicable and the leading order result is:
\begin{equation}
\frac{d}{dt}N_{p}=-2\left(  \frac{eE}{\hbar}\right)  ^{2}t\sum_{\mathbf{k}%
}\left[  \frac{h\partial_{p_{y}}h^{^{\ast}}-cc}{\varepsilon^{2}}\sin\left(
\frac{2\varepsilon t}{\hbar}\right)  \right]  ^{2}. \label{N_pert}%
\end{equation}
This is analogous to "linear response" for current. Immediately after the
switching on of electric field (times of order $t_{\gamma}$) the behaves as
$t^{3}$. For $t_{\gamma}<t<t_{nl}$ the pair creation rate per unit area rises
linearly (with logarithmic corrections),
\begin{equation}
\frac{d}{dt}N_{p}\simeq\frac{2}{\pi}\left(  \frac{eE}{\hbar}\right)  ^{2}%
t\log\left(  \frac{t}{t_{\gamma}}\right)  , \label{N_lin}%
\end{equation}
and is dominated by the neighborhood of the Dirac points.

However it is clear from Fig.1 that the expansion breaks down at $t_{nl}$,
when the rate stabilizes approximately at%

\[
\frac{d}{dt}N_{p}=3.7v_{g}^{-1/2}\left(  \frac{eE}{\hslash}\right)  ^{3/2}.
\]
and scales as the power $E^{3/2}$. The rate continues to rise in a series of
small jumps till Bloch oscillations set in. At that stage (actually at about
$t_{B}/4$) number of electrons elevated into the conduction band becomes of
order one, consistent with Eq.(\ref{t_Bloch}). Then it oscillates. The power
$E^{3/2}$ is, up to a constant, the same as the rate of the vacuum breakdown
due to the pair production calculated beyond perturbation theory by Schwinger
in the context of particle physics (when generalized to the 2+1 dimensions and
zero fermion mass \cite{Schwinger,Cohen}). This is not surprising since the
power $E^{3/2}$ is dictated by dimensionality assuming ultra - relativistic
approximation is valid. However the physical meaning is somewhat different. We
have used here a definition of the pairs number with respect to Fermi level of
the system before the electric field is switched on (equivalently when an
electrons are injected into a graphene sheet from a lead). This is different
not only from the Schwinger's path integral definition in which the Fermi
level is "updated" along the work of electric field and from the definition
proposed recently \cite{Cohen} in connection with graphene. The asymptotics at
very large times is not relevant for experimentally achievable ballistic
times, so that the predicted relatively short plateau segments are more important.

\section{Summary}

Ballistic transport in single graphene sheet near Dirac point was investigated
using the dynamic approach beyond linear response theory. We found that, while
the observation of the Bloch oscillations is difficult, there exists a novel
time scale $t_{nl}$, see Eq.(\ref{t_cros}), of transition to a nonlinear
regime which is within reach of current experimental techniques. The physics
of the ballistic transport in graphene can be described as a succession of
four time segments with different character.

(i) At microscopic ballistic times $t\sim t_{\gamma}$ the current reacts fast
to electric field and depends on microscopic details.

(ii) The current density at zero temperature stays constant $\sigma_{2}E$ for
ballistic times $t_{\gamma}<t<$ $t_{nl}$ and physics is partially universal in
the following sense. There are generally two contributions to the current.
While one contribution is dominated by Dirac points, the other is related to
the band structure. However the second contribution vanishes due to symmetry
properties of the Brillouin zone, see ref. \cite{Lewkowicz}.

(iii) For $t_{nl}<t<$ $t_{B}$ the current density during the flight would rise
above this value. It is dominated solely by the close vicinity of each of the
two Dirac points. Perhaps the increase of conductivity might be at least
partly responsible for the "missing $\pi$" problem \cite{Novoselov05,Castro},
namely that experimentally measured minimal conductivity is higher than
$\sigma_{2}$ even in suspended samples \cite{Andrei08}.

(iv) Finally at $t\sim$ $t_{B}$ Bloch oscillations set in. The physics is
again dominated again by the band structure, is "non - relativistic" and is
not directly related to the Dirac points.

It should be noted that in addition to limitations of the tight binding model
used which ignores impurities, interactions, deviation of the chemical
potential from the Dirac point and temperature beyond linear response such
"relativistic" effects like the pair annihilation neglected. For very large
electric fields the effects of radiation of energy into space (radiative
friction) might in principle be observable and should be investigated. On the
other hand influence of temperature and nonzero chemical potential in
nonlinear regime are expected to be similar to those in linear response
studied in \cite{Lewkowicz}.

\acknowledgments We are grateful to S. Gurvitz, E. Andrei, A. Morpurgo, J.
Goltzer, W.B. Jian for discussions. Work of BR and YK was supported by NSC of
R.O.C. grant \#972112M009048 and MOE ATU program. BR acknowledges the
hospitality of the Applied Physics Department of AUCS; M.L. acknowledges the
hospitality and support at Physics Department of NCTU.


\begin{thebibliography}{99}                                                                                               %


\bibitem {GeimPRL08}S.V. Morozov et al, Phys. Rev. Lett, \textbf{100}, 016602 (2008).

\bibitem {Andrei08}X. Du, I. Skachko, A. Barker and E. Y. Andrei, Nature
Nanotechnology \textbf{3}, 491 (2008).

\bibitem {Davis}J. H. Davis, "The physics of low dimensional semiconductors",
Cambridge University Press (1998).

\bibitem {Castro}A. H. Castro Neto et al, Rev. Mod. Phys. 81, 109 (2009).

\bibitem {Schwinger}J. Schwinger, Phys. Rev. \textbf{82}, 664 (1951).

\bibitem {Ziegler}K. Ziegler, Phys. Rev. \textbf{B75}, \ 233407 (2007).

\bibitem {Lewkowicz}M. Lewkowicz and B. Rosenstein, Phys. Rev. Lett,
\textbf{102}, 106802 (2009).

\bibitem {Fradkin}E. Fradkin, Phys. Rev. \textbf{B 33}, \ 3257 (1986); P.A.
Lee, Phys. Rev. Lett, \textbf{71}, 1887 (1993); V. P. Gusynin and S. G.
Sharapov, Phys. Rev. \textbf{B} \textbf{73}, \ 245411 (2006).

\bibitem {Katsnelson06}M.I. Katsnelson, Eur. Phys. J. \textbf{B 51}, 157
(2006); J. Tworzydlo et al, Phys. Rev. Lett. \textbf{96}, 246802 (2006).

\bibitem {Cohen}A. Casher, H. Neuberger and S. Nussinov, Phys. Rev. \textbf{D}
\textbf{20},\ 179 (1979); D. Allor, T. D. Cohen and D.A. McGady, Phys. Rev.
\textbf{D} \textbf{78}, \ 096009, (2008).

\bibitem {Gitman}E.S. Fradkin, D.M. Gitman and S.M. Shvartsman,
\textit{Quantum Electrodynamics with Unstable Vacuum} (Springer-Verlag, Berlin 1991).

\bibitem {Chicone}C. Chicone, "Ordinary differential equations with
applications", Springer - Verlag, New York (1999).

\bibitem {Novoselov05}K.S. Novoselov et al, Nature \textbf{438}, 197 (2005);
Y. Zhang et al, Nature \textbf{438}, 201 (2005).
\end{thebibliography}
\end{document}